\documentstyle[12pt]{article}
\begin{document}
\title{Fermi Temperature Magnetic Effects}
\author{B.G. Sidharth$^*$\\
Centre for Applicable Mathematics \& Computer Sciences\\
B.M. Birla Science Centre, Adarsh Nagar, Hyderabad - 500 063 (India)}
\date{}
\maketitle
\footnotetext{$^*$Email:birlasc@hd1.vsnl.net.in; birlard@ap.nic.in}
\begin{abstract}
Recent results that an assembly of Fermions below the Fermi temperature
would exhibit anomalous semionic behaviour are examined in the context
of associated magnetic fields.
\end{abstract}
Recently it was shown that \cite{r1} below the Fermi temperature
Fermions exhibit an anomalous Bosonization: They obey statistics
in between the Fermi-Dirac and Bose-Einstein.\\
In general given an assembly of $N$ Fermions, if $N_+$ is the average
number of particles with spin up, the magnetisation is given by \cite{r2},
\begin{equation}
m = \mu (2N_+ - N),\label{e1}
\end{equation}
where $\mu$ is the electron magnetic moment. In the usual theory, $N_+ \approx
\frac{N}{2}$ so that $m$ given in (\ref{e1}) is small. However semionic
statistics implies
\begin{equation}
N_+ = \beta N, \frac{1}{2} < \beta < 1,\label{e2}
\end{equation}
As N is generally very large, infact the number of particles is $\sim 10^{23}$ per cc or more,
the use of inequality (\ref{e2}) in (\ref{e1}) can give appreciable values
for $m$.\\
In other words given such an assembly of Fermions, the introduction of, for
example, an uniform magnetic field $B$ would lead to an energy creation
$\sim mB$, where initially the Fermion assemhbly had negligible magnetism.\\
Moreover the semionic behaviour could result in magnetic reversals, if
for example the external field $B$ changed its direction.\\
The relevance of the above considerations is varied. It must be mentioned
that in different conditions, the Fermi temperature of the assembly itself
would have a very wide spectrum starting from small values. For example, in Neutron stars the Fermi
temperature $\sim 10^7K$, while in the solid core of the earth it is $\sim 10^4$.
In these cases $N \sim 10^{58}$ and $10^{48}$ respectively. Indeed in both
these cases the prevalent magnetic field follows from (\ref{e1}) \cite{r3}.\\
Similarly the magnetism of the planet Jupiter can also be explained by
(\ref{e1}).\\
In the case of the earth there are magnetic reversals which are usually
attributed to the liquid core activity of the earth, though there is no
convincing explanation. Interestingly the Mars Global Surveyor space craft
detected such magnetic reversals on Mars also, in the last week of April,
1999. Mars has no liquid core and tectonic activity so that, in conventional
theory, this would point to a much earlier epoch of such possible activity.
However this is not required in the scenario presented above.


\begin{thebibliography}{99}
\bibitem {r1} Sidharth, B.G., 1999, Anomalous Fermions,  Journal of
Statistical Physics 95, 3/4.
\bibitem {r2} Huang, K., 1975, Statistical Mechanics, Wiley Eastern, New
Delhi.
\bibitem {r3} Sidharth, B.G., 1999, Magnetism of Neutron Stars and Planets,
xxx.lanl.gov/Physics/9904059.
\end{thebibliography}
\end{document}